# Modeling and Design of a Diaphragm-Integrated Photonic Waveguide Microphone

**Xiaoyu Niu, Yuqi Meng, Zihuan Liu, Ehsan Vatankhah, Neal Hall**
Department of Electrical and Computer Engineering, The University of Texas at Austin, Austin, Texas, 78712, USA; xyniu@utexas.edu; yuqimeng@utexas.edu; zihuanliu@utexas.edu; e.vatankhah@utexas.edu; nahall@utexas.edu

## I.ABSTRACT

Can an integrated optical readout improve the useful operating range of a miniature microphone? We address this question with an analytical model of a diaphragm carrying the sensing arm of a Mach–Zehnder interferometer (MZI). The model links pressure-induced bending to waveguide elongation and then to photocurrent, allowing mechanical sensitivity and optical detection limits to be assessed separately. Optical shot noise and back-cavity thermomechanical noise set the pressure resolution, while a 10% total harmonic distortion criterion defines acoustic overload. Two operating regimes are examined: a compact MEMS microphone with 1 mW optical input and a measurement microphone with 100 mW optical input. For the compact case, a 1 mm diameter, 1 μm thick, stress-free silicon diaphragm and 100% modulation efficiency give a predicted signal-to-noise ratio of 78 dB and acoustic overload pressure of 124 dB. The higher-power case lowers the modeled noise floor and increases dynamic range relative to the selected measurement microphone benchmark. These estimates rely on favorable assumptions and do not establish a substantial overall advantage in conventional microphone applications. Remote optical interrogation offers a separate motivation for harsh-environment sensing, but the temperature tolerance of the proposed sensing head remains unverified. The analysis identifies the design tradeoffs and measurements needed to test the predicted performance.

Keywords:

Integrated photonic sensing; Diaphragm strain; Mach–Zehnder interferometry; Optical MEMS microphone; Acoustic sensitivity; Shot noise; Acoustic overload

## II.RESEARCH QUESTION AND RELATION TO PRIOR WORK

This study asks whether removing the electrical diaphragm interface and a separate backplate yields a useful performance advantage after optical noise and overload are included. Capacitive[9] and piezoelectric[10] microphones provide the principal electrical comparisons; previous integrated-waveguide microphones establish the feasibility of the optical sensing approach.[11–13] The diaphragm-integrated MZI concept was introduced in our 2023 meeting abstract.[22] Here, we examine that concept through a coupled mechanical and optical model, a pressure-referred noise budget, and two application-specific design calculations. The scope is a theoretical assessment under stated assumptions. The calculations do not constitute experimental validation of a fabricated waveguide microphone, and the comparisons use the particular historical devices identified below.

Optical phase measurements can resolve small changes in a sensing path.[1] The Michelson–Morley experiment[2] and LIGO[3] provide historical and large-scale examples, respectively, as illustrated in Figure 1. Integrated photonics brings related techniques to biomedical sensing,[4] rotation measurements,[5,6] chemical detection,[7] and accelerometry.[8] Our cantilever MZI and heterodyne silicon-nitride accelerometers use strain in a waveguide-bearing structure to modulate optical phase.[20,21] Their relevance to the microphone is this mechanical-to-optical coupling, rather than a direct transfer of their measured performance.

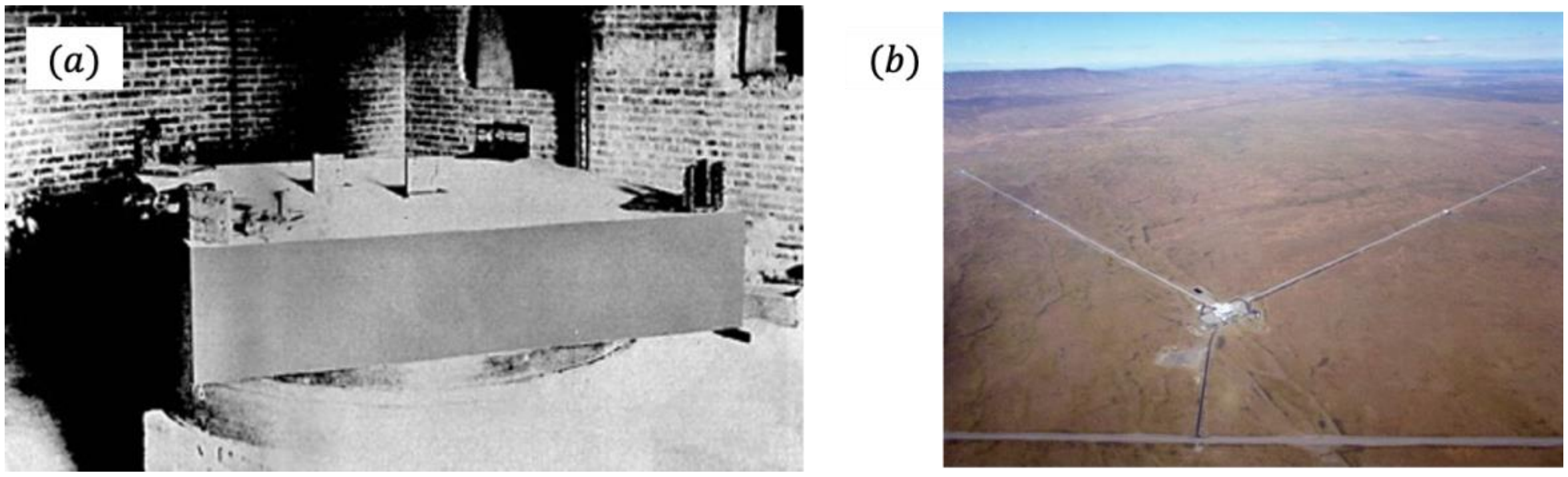

***Figure 1 Examples of interferometric measurement systems. (a) The 1887 Michelson–Morley apparatus, supported by a stone slab floating in an annular mercury trough. (b) The LIGO installation near Hanford, Washington, USA.***

The physical sensing path distinguishes this microphone from related acoustic devices. Our lithium niobate microphone uses piezoelectric detection,[23] our electrostatic device operates a MEMS microphone architecture as an ultrasound transmitter,[24] and our free-space LDV method detects sound-induced changes in refractive index.[25] In the present model, diaphragm strain changes the length of a guided optical path, so the waveguide and diaphragm must be considered together.

# III.NOISE SOURCES AND TRANSDUCTION CHOICES

Removing a sensing interface is useful only if the resulting noise budget improves. For the capacitive microphone summarized in Figure 2, backplate-related noise, back-volume noise, and ASIC noise are major contributions.[14] Their relative sizes and lumped-element representation appear in Figure 2(a) and Figure 2(b), respectively. Lithium niobate and packaged AlN microphone studies likewise show that mechanical sensitivity and electrical readout must be evaluated together.[23,26] We use this framework to identify which contributions the optical architecture removes and which remain.

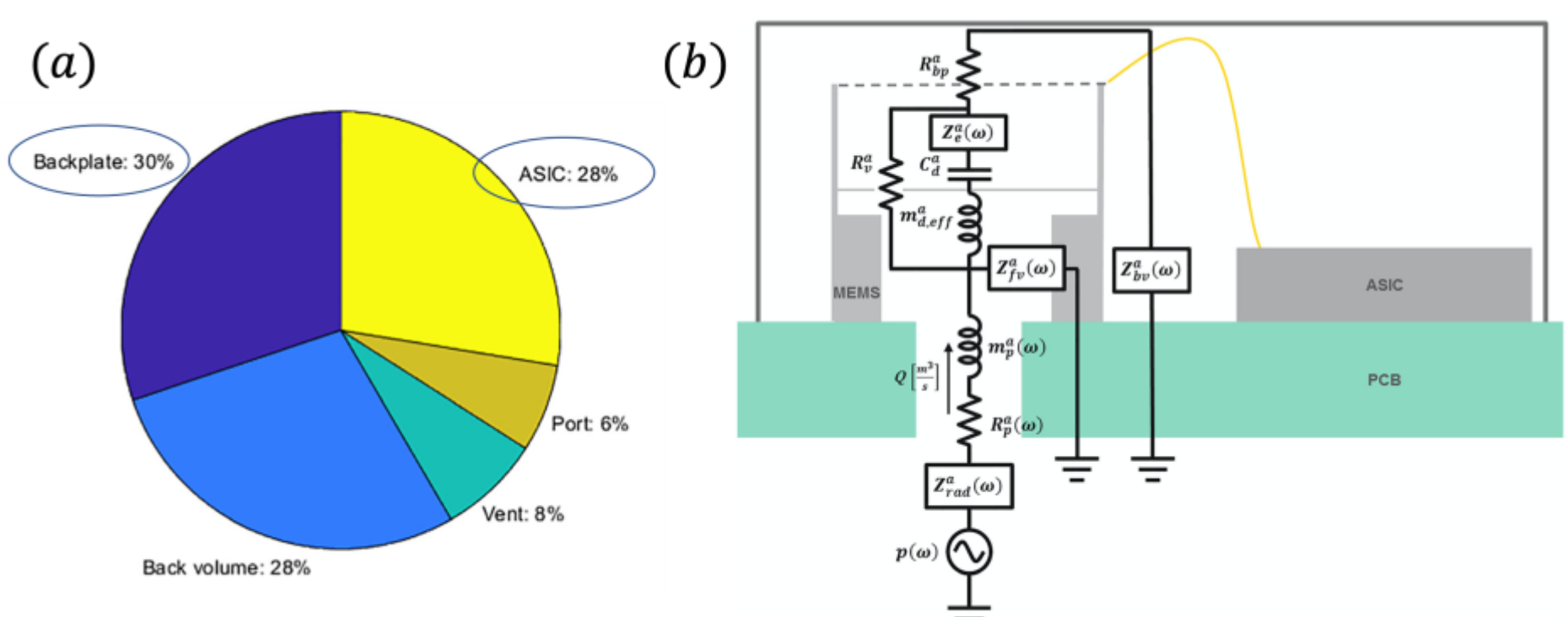

***Figure 2 Noise budget and equivalent model for a capacitive MEMS microphone. (a) Relative noise contributions. (b) Lumped-element representation of the sensor.***

The optical-grating design in Figure 3(a) supplies an instructive comparison.[15] Its MEMS diaphragm, optoelectronics, ASIC, lid, and PCB form a complete microphone package. In Figure 3(b), a VCSEL illuminates a grating; reflections from that grating and the underlying silicon diaphragm interfere at a photodiode. Motion changes their relative phase and therefore the detected intensity. Larger grating openings reduce the thermomechanical contribution associated with a capacitive backplate, while optical detection avoids the high-input-impedance interface represented by the ASIC contribution in Figure 2(a). The PCB-mounted prototype is shown in Figure 3(c). Its $4.14\ mm \times 4.15\ mm \times 1.38\ mm$ package achieved a reported noise floor of 22.00 $dBA$ and an SNR of 72 $dB$. The comparison in Figure 3(d) concerns microphones available in 2014, when the prototype in Figure 3[15] was developed. It should therefore be read as a historical benchmark rather than a comparison with every capacitive microphone currently available.

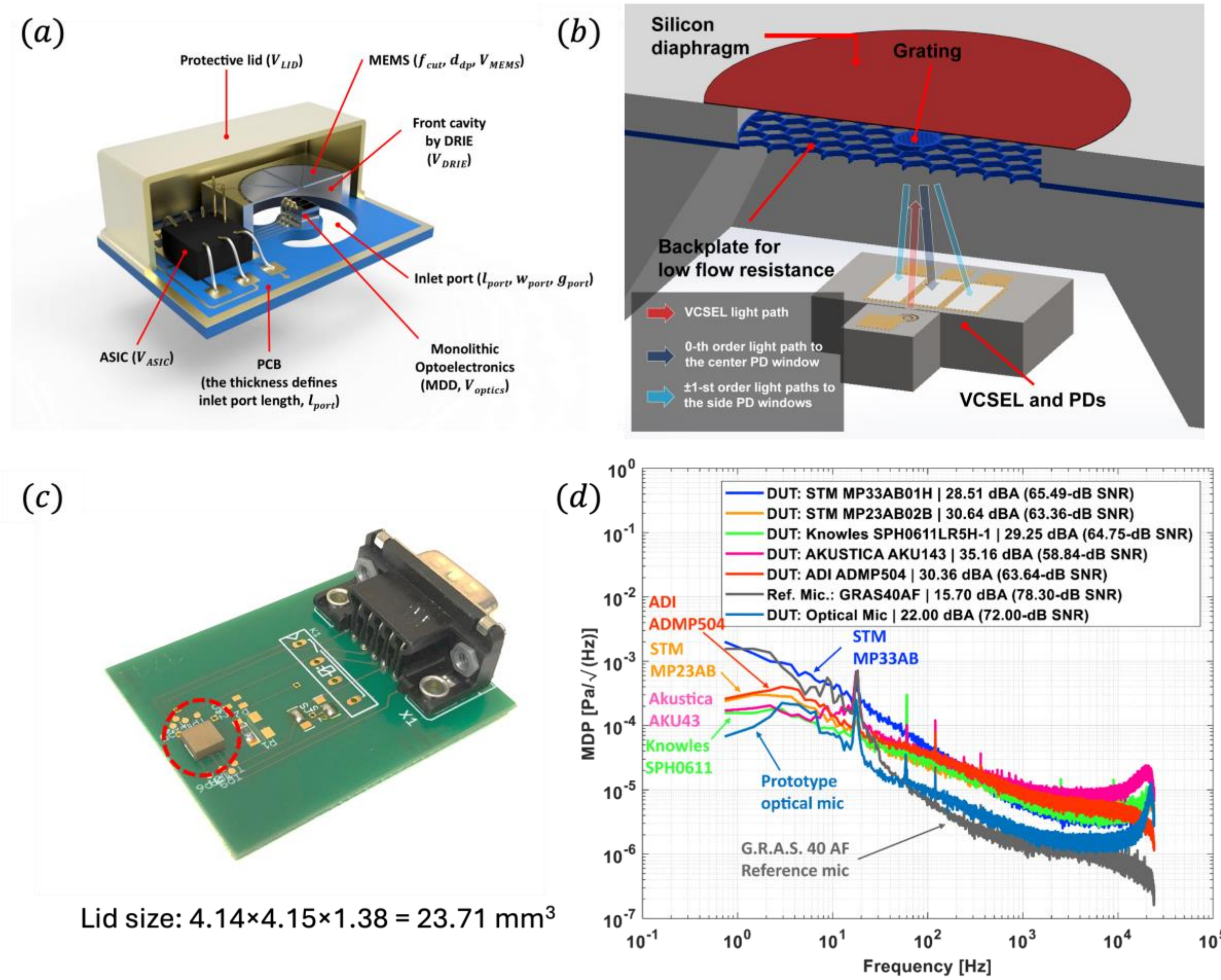

***Figure 3 Optical-grating microphone and historical performance comparison. (a) CAD view of the assembly. (b) Detail of the MEMS die and optical components. (c) Prototype mounted on a test PCB. (d) MDP relative to commercial MEMS microphones available in 2014.***

The architectural comparison in Figure 4 separates two changes to the capacitive noise budget. First, the capacitive backplate in Figure 4(a) is replaced by the more open optical grating in Figure 4(b), reducing the backplate-related thermomechanical contribution represented by the purple sector in Figure 4(b). Second, the integrated-waveguide design in Figure 4(c) dispenses with that opposing structure entirely. Both optical approaches avoid the high-input-impedance diaphragm interface. The diagrams consequently omit the noise associated with that interface; they do not represent a complete noise model of the photodetector and downstream electronics.

Our low-output-impedance magnetoelastic vibration sensors provide another example of how transduction changes the need for local electronics.[31] For a waveguide microphone, this architectural benefit must be assessed separately from detector noise, electronic noise, and shot noise.

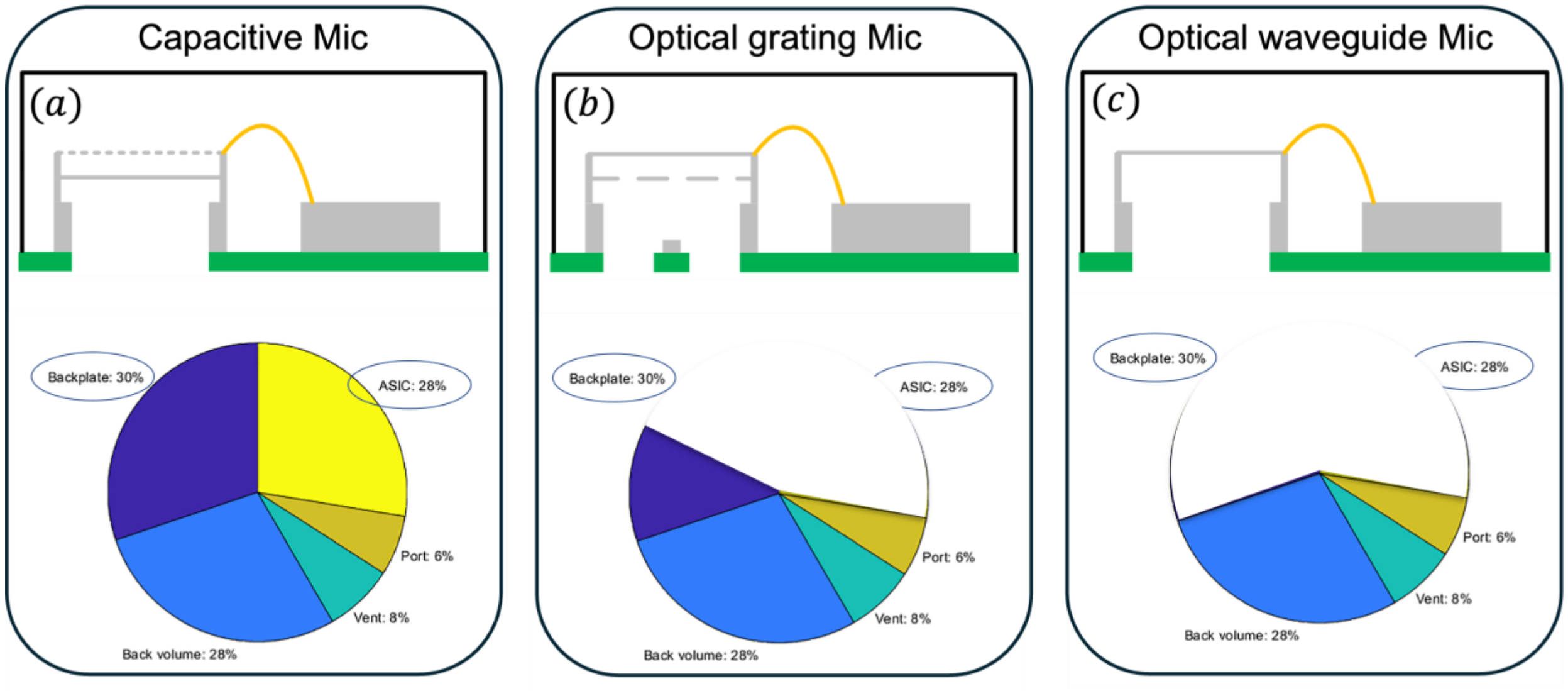

***Figure 4 Comparison of capacitive, optical-grating, and integrated-waveguide MEMS microphone architectures with their associated conceptual noise budgets.***

# IV.DEVICE MODEL AND PERFORMANCE DEFINITIONS

The configuration in Figure 5 uses a diaphragm-only MEMS die with conventional packaging components, including a PCB, ASIC, bonding wires, and lid. The absence of a backplate is apparent in Figure 5(a). The MZI in Figure 5(b) places its reference arm on the supporting substrate and its sensing arm on the diaphragm. MMI components divide and recombine the guided light. Diaphragm bending changes the sensing-arm path length; interference with the reference arm converts the resulting phase change to an intensity change. A representative waveguide implementation is shown in Figure 5(c).[12] A diaphragm-embedded microresonator with thermo-optic tuning has also been explored by our group.[27] That related architecture uses a different optical transfer function and is outside the MZI model developed here.

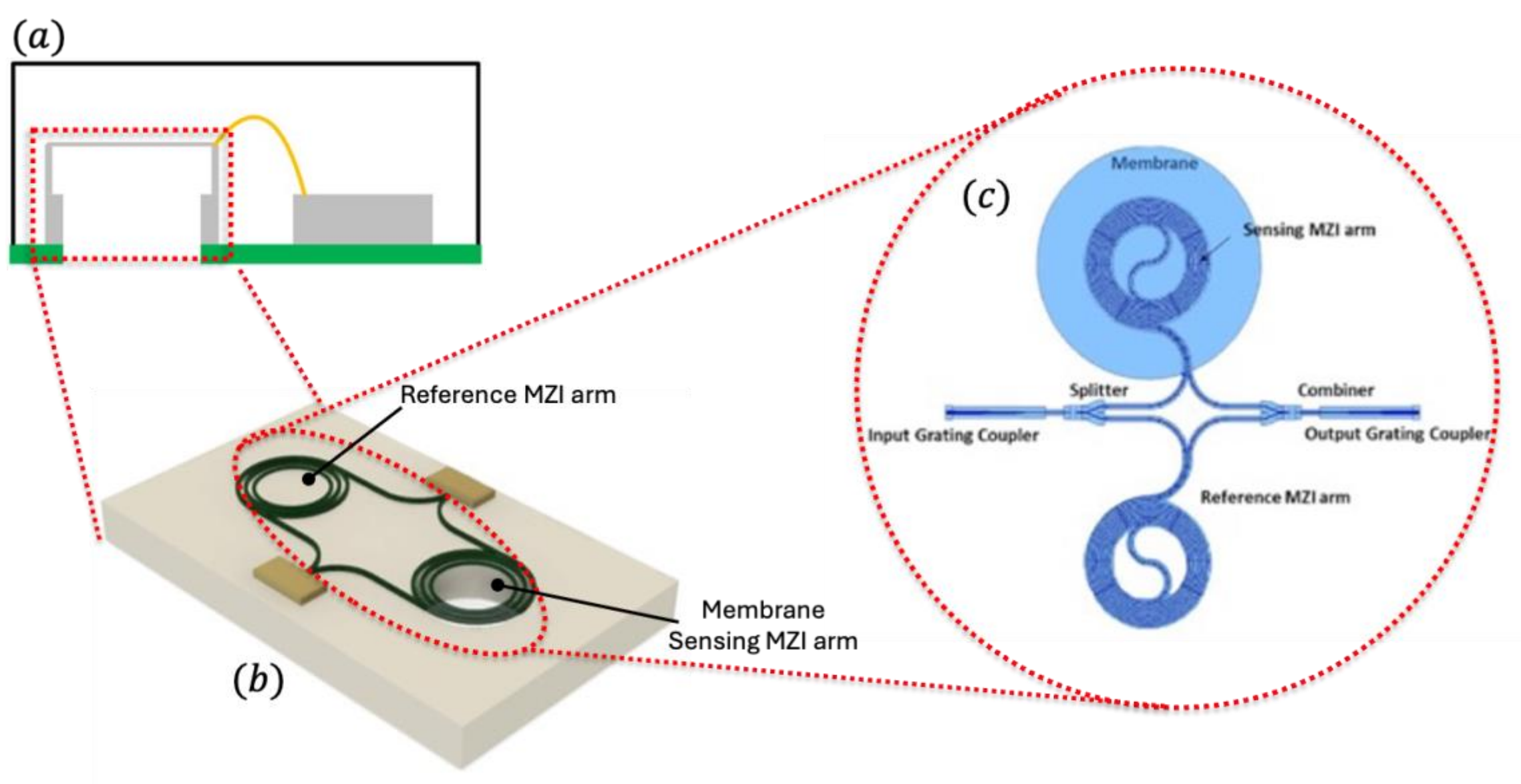

***Figure 5 Waveguide microphone layout. (a) Proposed package arrangement. (b) Reference arm on the supporting die and sensing arm on the diaphragm. (c) Representative MZI waveguide implementation.***

The integration concept combines waveguides, a laser,[16] and a photodetector[17] on one die (Figure 6). An edge-emitting source is preferred for the proposed heterogeneous arrangement over a VECSEL. Alternatively, external optical components can interrogate the sensing die, as illustrated by our photonic accelerometer implementations.[20,21] This distinction matters when evaluating packaging constraints and the possibility of locating electronics away from the sensing head.

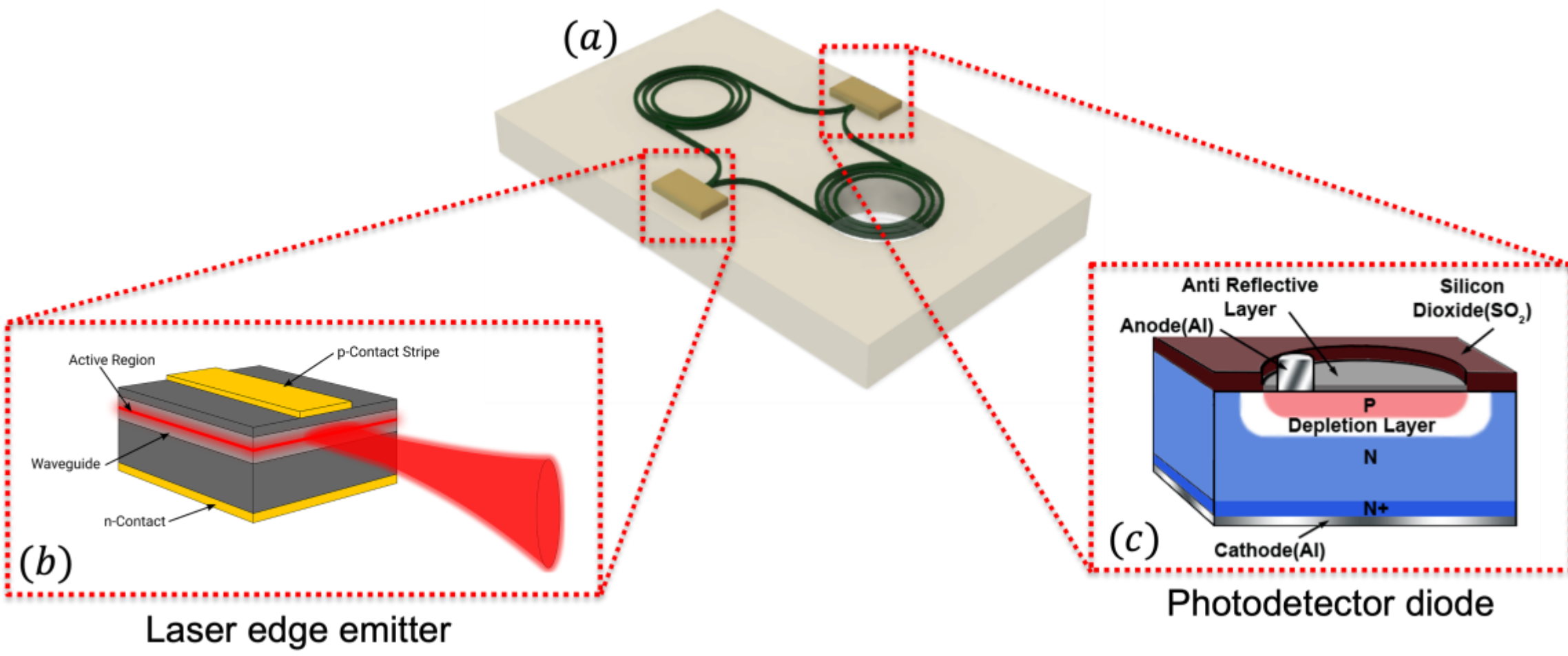

***Figure 6 Proposed integration of optical components on the MEMS die. (a) Overall die concept. (b) Laser-source placement, with an edge-emitting source preferred for the proposed arrangement. (c) Photodetector placement.***

## A.OPTICAL RESPONSE AND INTRINSIC DYNAMIC RANGE

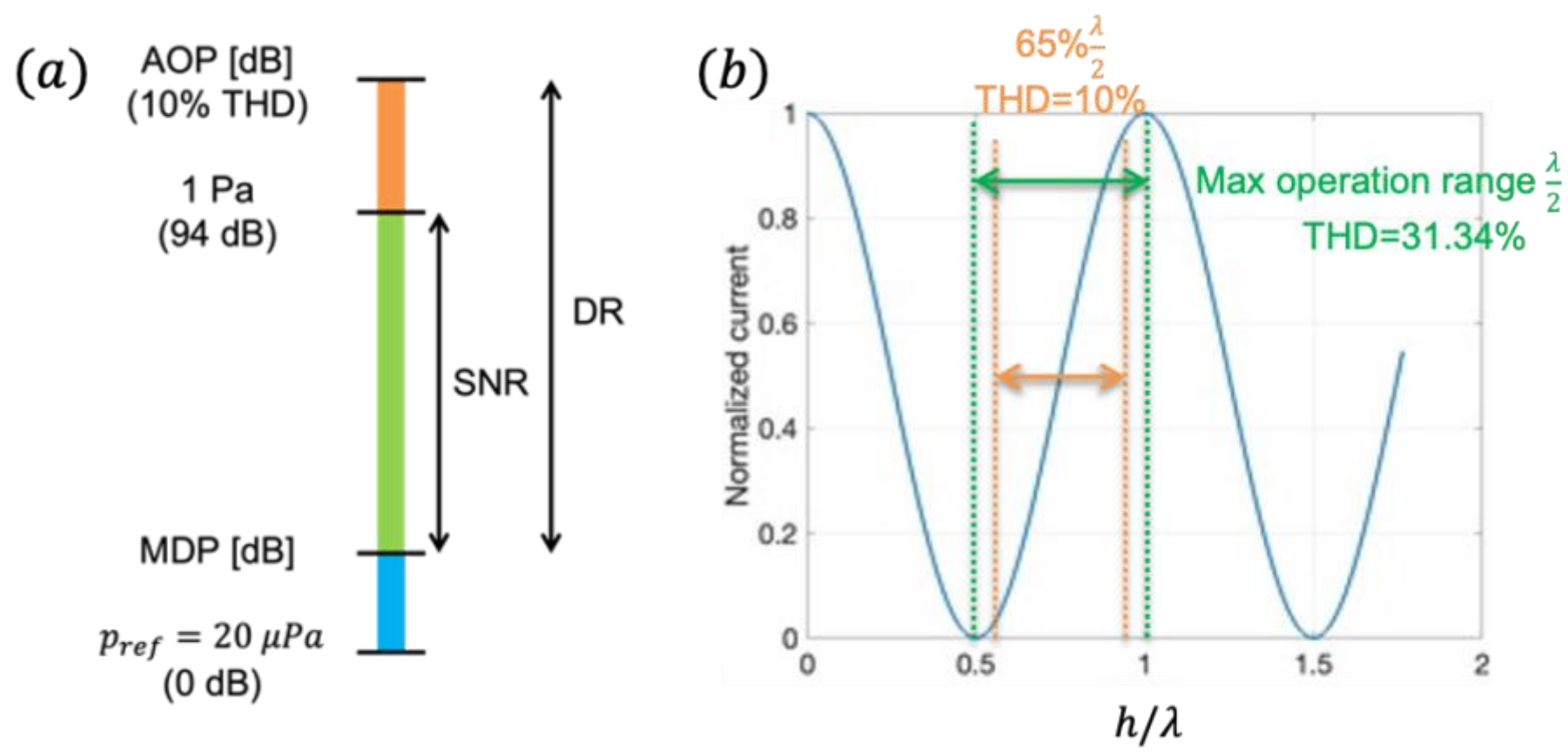

***Figure 7 Performance metrics and interferometric response. (a) Acoustic-pressure scale relating MDP, SNR, DR, and AOP. (b) Photodetector current as a function of normalized optical path difference.***

The pressure range summarized in Figure 7(a) is bounded by MDP at the low end and AOP at the high end. SNR uses a specified reference pressure, while DR relates the upper signal boundary to the noise floor. We first isolate the optical contribution by assuming a linear mechanical response and direct intensity detection about a fixed bias point. The interferometric nonlinearity then determines the admissible optical excursion. This bound does not apply unchanged to other readout methods: heterodyne in-phase and quadrature detection can extend the phase range, as in our silicon-nitride accelerometer.[21] Such a microphone implementation would need its own noise and overload analysis.

For the MMI output illustrated in Figure 5(c), the optical-power transfer function is

$$L = \frac{L_0}{2}\left[1 + \cos\left(\frac{2\pi h}{\lambda}\right)\right]\ [W] \quad (1)$$

Here, $L_0$ denotes the input optical-power parameter for detection at the photodiode in Figure 6(c), $h$ is the path-length difference between the MZI arms, and $\lambda$ denotes the optical wavelength. Conversion of the optical output to photodiode current gives

$$i = R_{op} \cdot ME \cdot L\ [A] \quad (2)$$

The photodetector responsivity $R_{op}$ is expressed in $\left[\frac{A}{W}\right]$, and the modulation efficiency $ME$ in [%]. Combining Eqs. (1) and (2) gives the current-versus-path-difference curve in Figure 7(b). The green dashed limits mark the maximum operating interval of $\frac{\lambda}{2}$. We define overload at 10% total harmonic distortion (THD), restricting the allowed optical path excursion to $65\% \frac{\lambda}{2}$. This distortion criterion supplies the upper signal boundary used to calculate DR.

The maximum slope of the photocurrent transfer function defines the optical sensitivity:

$$S_{op} = \frac{di}{dh}\Big|_{max} = \frac{\pi}{\lambda}\sqrt{ME \cdot R_{op} L_0}\ \left[\frac{A}{m}\right] \quad (3)$$

An ideal optical noise bound is obtained by retaining shot noise alone. Dividing its current amplitude spectral density by the optical sensitivity gives the minimum detectable displacement (MDD):

$$MDD = \frac{i_{shot}}{S_{op}} = \frac{\sqrt{qR_{op}L_0}}{\frac{\pi}{\lambda}\sqrt{ME \cdot R_{op}L_0}} = \frac{\lambda}{\pi}\sqrt{\frac{q}{ME \cdot R_{op}L_0}}\ \left[\frac{m}{\sqrt{Hz}}\right] \quad (4)$$

MDD refers here to a resolvable difference in optical path length at the readout stage. MDP instead expresses the pressure resolution of the complete microphone, including mechanical conversion. Using the distortion-limited path excursion and MDD, the intrinsic optical dynamic range becomes

$$DR = \frac{65\% \cdot \frac{\lambda}{2}}{MDD} = 65\% \cdot \frac{\pi}{2} \cdot \sqrt{\frac{ME \cdot R_{op}L_0}{q}} \quad (5)$$

Pressure-referred noise permits comparison across transduction mechanisms. Our magnetostrictive hydrophone study estimates MDP from self-noise and sensitivity.[32] The same accounting for a waveguide microphone requires a stated bandwidth and frequency-dependent sensitivity, with calculated limits kept distinct from measured noise floors.

Expressed in $dB$, the intrinsic $DR$ follows from Eq. (5); in linear units it increases with the square root of optical input power. We evaluate the model for $ME = 50\%$ or $100\%$, with $R_{op} = 0.5\ \left[\frac{A}{W}\right]$ and $q = 1.6 \times 10^{-19}\ [C]$. The results in Table 1 list the calculated $DR$ alongside a state-of-the-art MEMS benchmark. That benchmark provides approximately 110 $dB$ of dynamic range at about 1 $mW$ of power consumption. Comparing optical input power with the power consumed by a complete microphone requires care: the calculated optical $DR$ alone does not establish a system-level advantage for either $ME = 100\%$ or $ME = 50\%$.

***Table 1 Intrinsic optical dynamic range calculated at three input powers and two modulation efficiencies, with a MEMS microphone benchmark.***

| $L_0$[mW] | $DR[dB]$ ($ME = 100\%$) | $DR[dB]$ ($ME = 50\%$) | SOA MEMS Mic |
|---|---|---|---|
| 1 | 115 | 112 | 110dB |
| 10 | 125 | 122 | |
| 100 | 135 | 132 | |

## B.PRESSURE SENSITIVITY AND ACOUSTIC OVERLOAD

To express SNR and overload in acoustic units, the optical response must be combined with diaphragm mechanics. Pressure first produces a path-length difference, which is then converted to photocurrent by the transfer function in Figure 7. For small mechanical deflections, the two sensitivities multiply. Consequently, the pressure-response curve in Figure 8 has the same interferometric form as Figure 7(b), with acoustic pressure replacing optical path difference on the horizontal axis. The blue curve describes the static response relative to atmospheric pressure. An acoustic signal produces oscillations around the bias point chosen at the steepest slope of that curve. Operation below the adopted AOP criterion confines this excursion to 65% of the full operating interval, indicated by the orange dashed boundaries.

A single sensitivity value does not establish bandwidth. For example, our Terfenol-D inertial sensor exhibits a jerk response and an acceleration-sensitivity slope of +6 dB/octave below resonance, supported by measurements and finite-element modeling.[33] That response is not a microphone prediction; it illustrates why frequency dependence must be checked for each sensing mechanism.

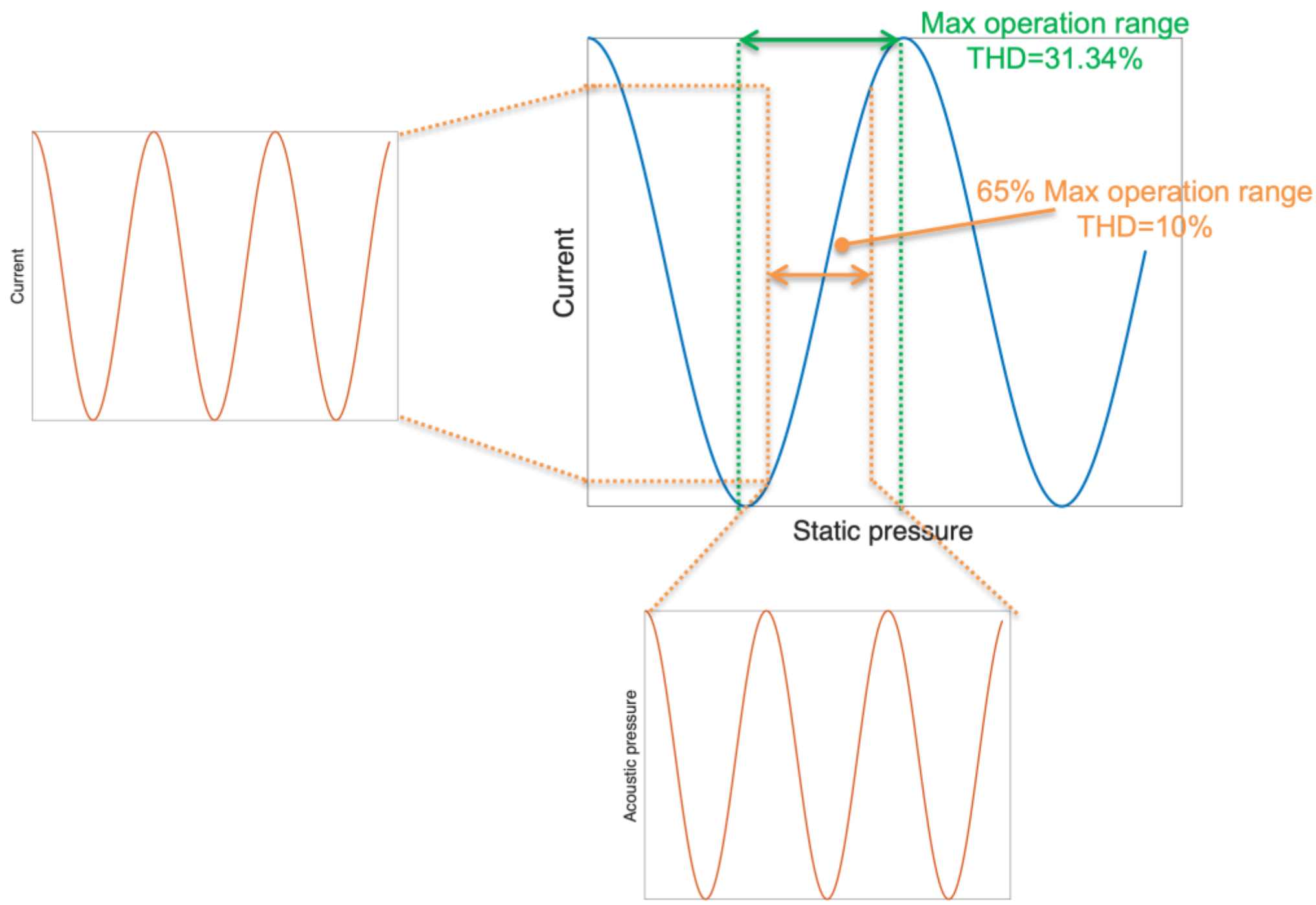

***Figure 8 Static pressure-to-photocurrent response, including the bias point and the pressure interval associated with the adopted distortion criterion.***

For the packaged configuration in Figure 5(a), the model retains two principal contributions to output-current noise: optical shot noise, $i_{shot}$, and back-cavity thermomechanical noise referred to the photocurrent, $i_{BC}$. Combining their mean-square contributions gives the SNR for a 1 Pa input:

$$SNR = \frac{1[Pa] \cdot S_T \left[\frac{A}{Pa}\right]}{\sqrt{i_{shot}^2 + i_{BC}^2}[A]} \quad (6)$$

$$S_T = S_{op}\left[\frac{A}{m}\right] \cdot S_m\left[\frac{m}{Pa}\right] \quad (7)$$

The overall current sensitivity to pressure, $S_T$, is the product of the optical sensitivity $S_{op}$ $\left[\frac{A}{m}\right]$ from Eq. (3) and the mechanical sensitivity $S_m$ $\left[\frac{m}{Pa}\right]$. The latter quantifies the path-length change produced by an incremental pressure:

$$S_m = \frac{dh}{dp}\left[\frac{m}{Pa}\right] \quad (8)$$

The applied acoustic pressure is denoted by $p$. Evaluating the derivative above requires the deformation-induced path difference $h$ to be related explicitly to $p$ through the diaphragm geometry.

We use the cross-section in Figure 9(a) to define the membrane radius $R$, membrane thickness $th$, waveguide offset $z$ from the membrane mid-plane, and radius $r_n$ of the $n^{th}$ waveguide turn. The cross-section shows a single turn for clarity; the actual sensing arm can contain several turns, as in Figure 5(c). Under pressure loading, the membrane bends as depicted in Figure 9(b). The displacements $w(r_n)$ and $\Delta r_n$ describe the vertical motion and radial change of the $n^{th}$ turn. Its radial expansion, $\Delta r_n$, increases the circumference of the $n^{th}$ turn. Summing these circumference changes yields the path-length change used in the model:

$$h = \sum_{n=1}^{N} 2\pi \cdot \Delta r_n \quad (9)$$

For $N$ turns, the radial change $\Delta r_n$ is linked to the deflection profile $w(r_n)$. The local rotation of the membrane shown in Figure 10 gives

$$\Delta r_n = -z \times \frac{dw(r_n)}{dr_n} \quad (10)$$

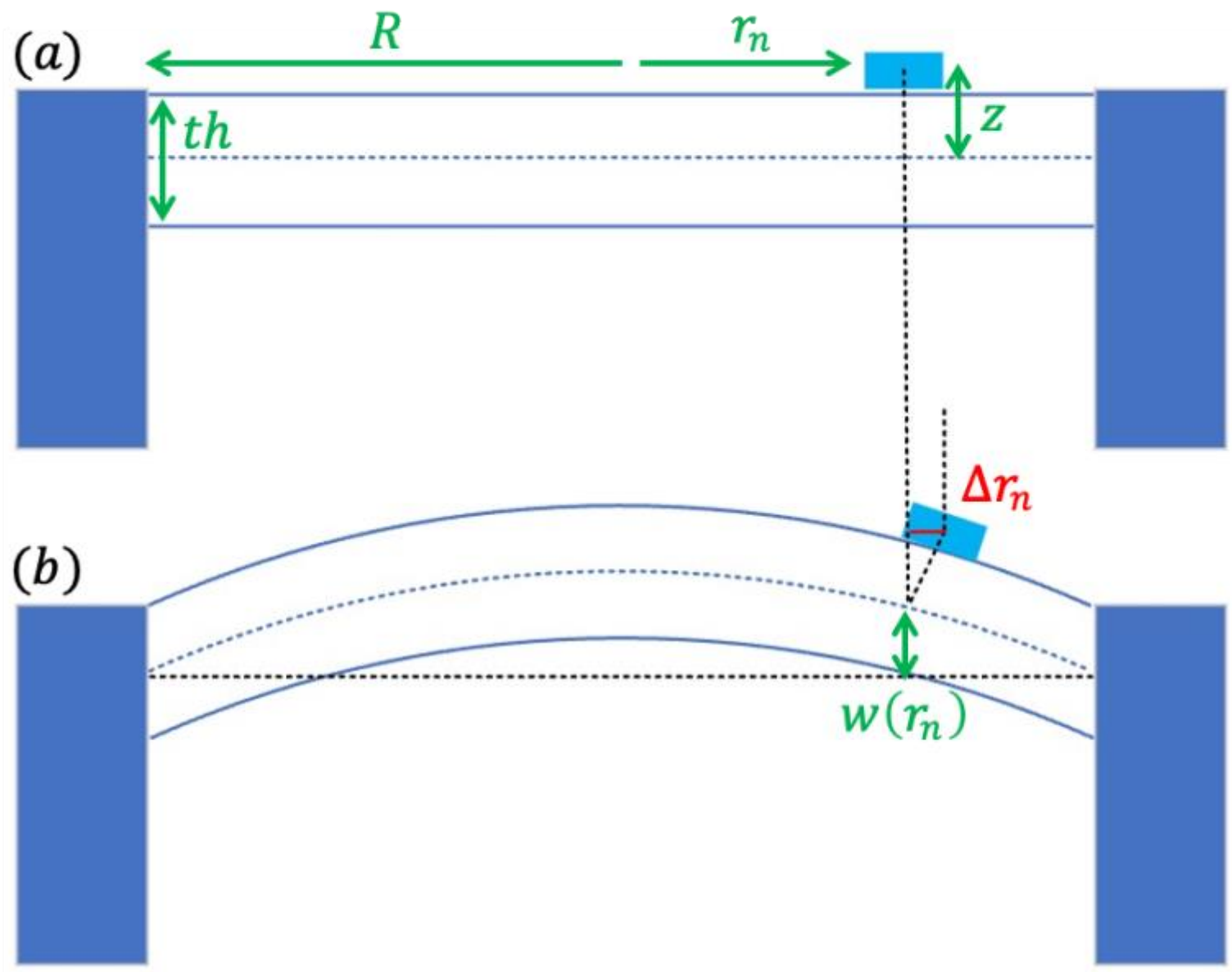

***Figure 9 Cross-sectional geometry of the waveguide sensing die. (a) Undeformed membrane and waveguide dimensions. (b) Pressure-induced bending and displacement of a waveguide turn.***

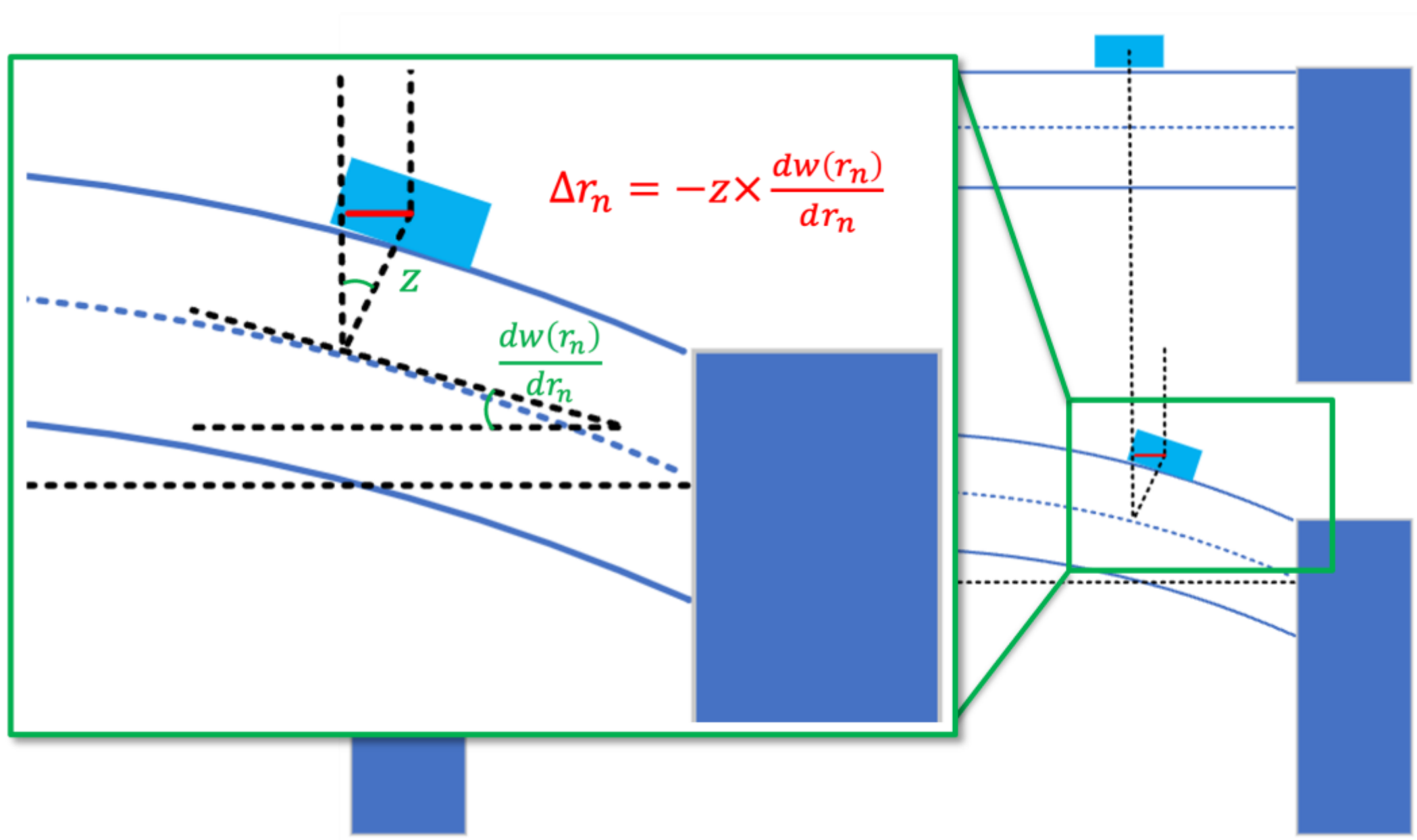

***Figure 10 Detail of the local membrane rotation and the resulting change in waveguide radius.***

We use a bending-dominated plate profile to evaluate a favorable mechanical limit. In the comparison considered here, intrinsic stress reduces the deflection and the associated waveguide elongation, so the plate-like case yields greater sensitivity. Turn count alone does not determine the result: each turn samples a particular part of the diaphragm strain field. Multiple serpentine loops increase sensitivity in our cantilever photonic accelerometer,[20] motivating joint consideration of optical-path length and placement when designing a diaphragm sensing arm.

The ideal model neglects the mechanical loading of the waveguide stack. Single-layer lithium niobate transducers using thickness-shear coupling[34] and bimorph lithium niobate microphones[23] provide alternative layer structures for comparison. A fabricated optical device would require its actual layer stack to be included in the mechanics.

$$w(r_n) = w_0\left(1-\frac{r_n^2}{R^2}\right)^2 \quad (11)$$

$$w_0 = C_m \cdot p \quad (12)$$

$$C_m = \left(\frac{16E\cdot th^3}{3R^4(1-\nu^2)} + \frac{4\sigma\cdot th}{R^2}\right)^{-1} \quad (13)$$

The center displacement is $w_0$, and $C_m$ denotes membrane compliance.[18] In the expression for $C_m$, the material parameters $E$, $\nu$, and $\sigma$ are Young's modulus, Poisson's ratio, and intrinsic stress, respectively.

The mechanical conversion can be tested independently of the optical readout. Our immersion lithium niobate PMUT study uses LDV and a hydrophone to relate motion and acoustic output.[35] For an airborne receiver, paired LDV and calibrated reference-microphone measurements would establish the pressure-to-diaphragm response under the same excitation.

Combining the geometric and deflection relations in Eqs. (10)–(12) with the path-length sum in Eq. (9) gives the pressure-induced path difference and its associated mechanical sensitivity:

$$h = \sum_{n=1}^{N}\frac{8\pi z r_n}{R^2}\left(1-\frac{r_n^2}{R^2}\right)C_m p\ [m] \quad (14)$$

$$S_m = \sum_{n=1}^{N}\frac{8\pi z r_n}{R^2}\left(1-\frac{r_n^2}{R^2}\right)C_m\ \left[\frac{m}{Pa}\right] \quad (15)$$

The optical sensitivity in Eq. (3), mechanical sensitivity in Eq. (15), and shot-noise term $i_{shot}$ can then be inserted into Eq. (6), giving

$$SNR = \frac{\frac{\pi}{\lambda}\sqrt{ME\cdot R_{op}L_0}\cdot\sum_{n=1}^{N}\frac{8\pi z r_n}{R^2}\left(1-\frac{r_n^2}{R^2}\right)C_m}{\sqrt{qR_{op}L_0 + i_{BC}^2}}$$

Converting the full optical path interval to pressure gives the corresponding operating interval:

$$Max\ operation\ range\ [Pa] = \frac{\frac{\lambda}{2}[m]}{S_m\left[\frac{m}{Pa}\right]} = \frac{\frac{\lambda}{2}}{\sum_{n=1}^{N}\frac{8\pi z r_n}{R^2}\left(1-\frac{r_n^2}{R^2}\right)C_m}[Pa] \quad (16)$$

Applying the 10% THD criterion to this interval yields the acoustic overload pressure:

$$AOP = 65\%\cdot Max\ operation\ range\ [Pa] \quad (17)$$

Measured overload also depends on the excitation and readout chain. Receiver nonlinearity encountered in our MEMS ultrasonic-transmitter study illustrates this issue.[36] Independent source and readout characterization is therefore needed to distinguish diaphragm distortion, optical distortion, and distortion generated elsewhere in a microphone measurement.

# V.APPLICATION CASES AND BENCHMARK COMPARISONS

The two cases differ in optical power and back-cavity noise. The compact consumer design assumes $L_0 = 1mW$; the measurement design allows $L_0 = 100\ mW$. Back-volume thermal noise becomes more restrictive as the package shrinks, as discussed by Kuntzman.[14] We assign the compact case a back-cavity EIN of 10 $dB$; this 10 $dB$ value is a modeling assumption for the selected high-performance MEMS regime. The measurement case omits that contribution. Both calculations neglect waveguide-induced changes to diaphragm mechanics. These choices define favorable comparison cases rather than specifications already achieved by a fabricated device.

Iterative optimization within these assumptions produces the representative MEMS and measurement designs shown in Figure 11 and Figure 12, respectively.

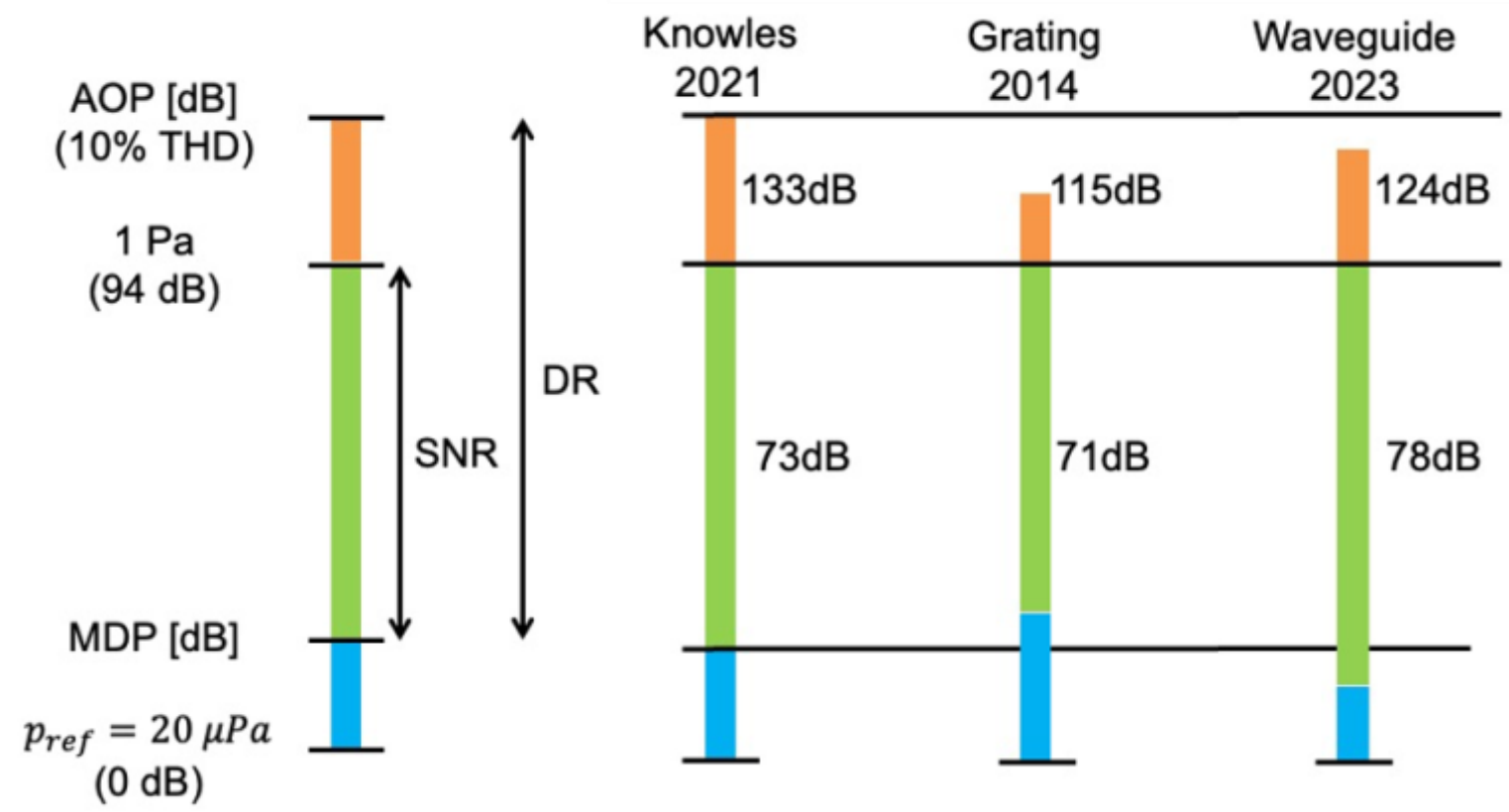

***Figure 11 Optimized compact waveguide microphone and comparison with the selected capacitive and optical-grating MEMS microphone benchmarks.***

The compact calculation uses a silicon diaphragm with diameter 1 $mm$, thickness 1 $\mu m$, and intrinsic stress 0 $MPa$. With 100% modulation efficiency, the model predicts an SNR of 78 $dB$ and an AOP of 124 $dB$. The comparison with the 2021 Knowles device[19] in Figure 11 gives similar DR, while comparison with the 2014 optical-grating prototype[15] favors the waveguide calculation in SNR and AOP. The practical conclusion depends on the entire microphone: these selected metrics do not establish an overall advantage once power, packaging, and implementation constraints are considered.

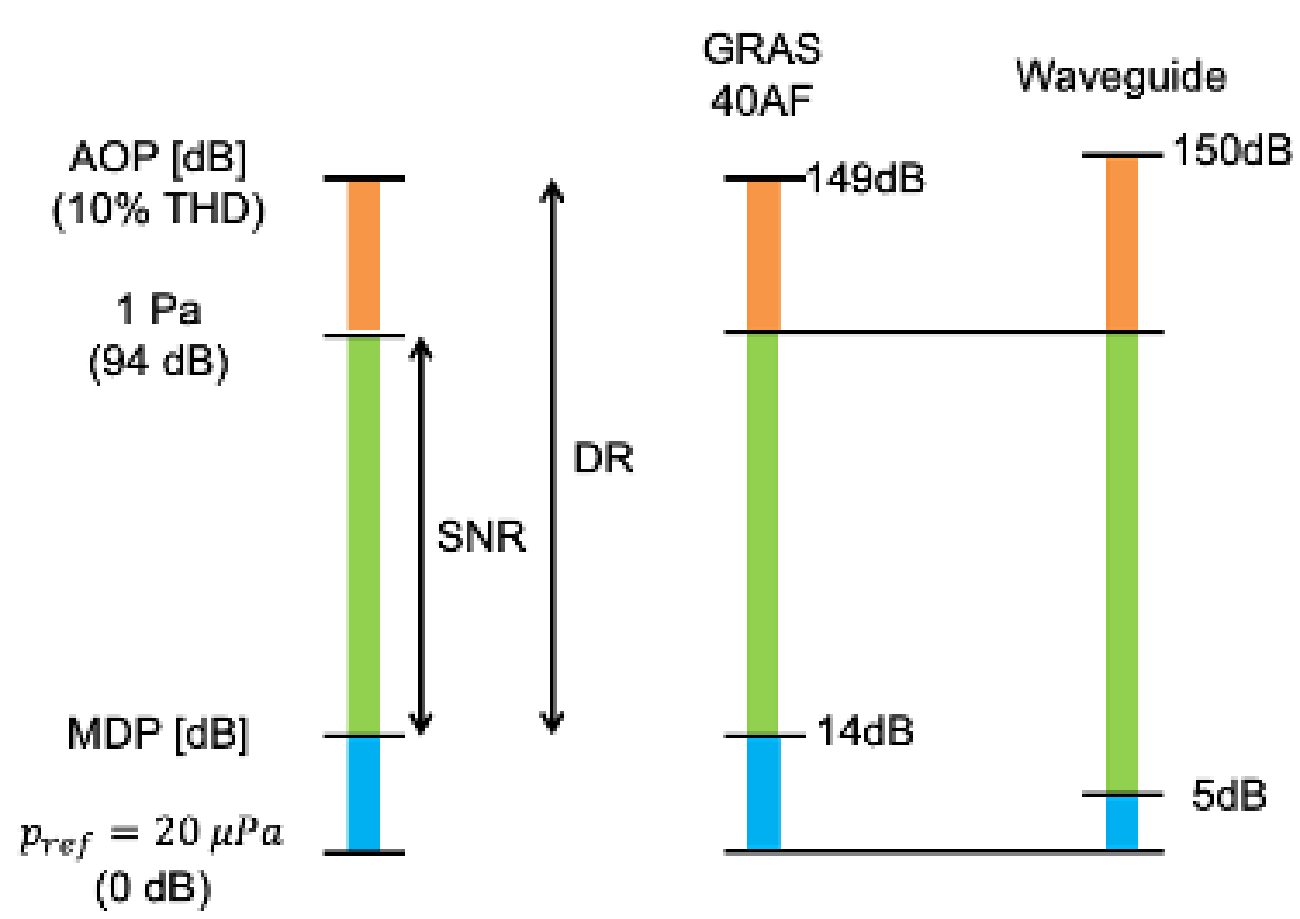

***Figure 12 Optimized waveguide design for precision measurement and comparison with the G.R.A.S. 40AF measurement microphone.***

At 100 $mW$ optical input with back-cavity noise omitted, the measurement case gives a lower modeled noise floor than the G.R.A.S. 40AF benchmark. The approximately 10 $dB$ increase in DR is driven mainly by MDP. Applications constrained by bandwidth or high-pressure waveform fidelity may gain less from that improvement. For context, our 300 kHz progressive-ultrasound measurements required at least 3 MHz bandwidth to resolve harmonics and shock formation.[28] No corresponding ultrasonic-band performance is established by the present audio-band calculation.

# VI.DISCUSSION

## A.INTERPRETATION AND LIMITS OF THE PREDICTIONS

The calculation assumes a dominant optical shot-noise term $i_{shot}$, negligible waveguide propagation attenuation, and a diaphragm whose response is governed by bending stiffness. These assumptions make the results useful as favorable performance estimates, but do not replace a full noise budget or a measured transfer function. Large displacement can also introduce mechanical nonlinearity. Strain stiffening treated in our state-space and LDV study of electrostatic transducers[29] is one reason to verify the small-deflection regime before using the microphone model at high acoustic pressure.

Time-domain and frequency-domain LDV measurements can identify amplitude-dependent behavior. Our AlN wedge-resonator study detects Duffing-type nonlinearity at large drive amplitudes.[37] Its geometry and frequency differ from those of the microphone, but the measurement approach is applicable to checking the assumed linear mechanical regime.

## B.DESIGN EXTENSIONS BEYOND THE PRESENT MODEL

The explored variables concern the membrane; waveguide loading and complete package optimization remain outside the calculation. Directionality would add another mechanical requirement. Our fly-inspired microphone combines omnidirectional pressure sensing with two orthogonal in-plane pressure-gradient responses on one diaphragm.[30] Coupling waveguides to those modes would require a new sensing layout rather than applying the present pressure model unchanged.

Field shaping could also be investigated separately from transducer design. Coupled Helmholtz-resonator gratings redirect acoustic energy over a range of frequencies and incidence angles.[38] Their demonstrated scale is macroscopic, so a miniature microphone application would need new scaling, loss, and packaging analysis.

## C.PROSPECTS FOR HIGH TEMPERATURE SENSING

A possible motivation beyond the conventional audio benchmarks is physical separation of the sensing die from its source and electronics. Fibers could connect a diaphragm-and-waveguide sensing head to a remotely located laser, detector, and ASIC. This arrangement makes high-temperature acoustic sensing worth evaluating, including for supersonic and geothermal measurements, but does not establish thermal robustness. The relevant operating limit would depend on the diaphragm stack, optical coupling, packaging, and stability of the interferometric bias.

Thermal operation and survival after exposure are distinct specifications. Stable operation to 600 °C and survival to 900 °C have been reported for our bimorph lithium niobate PMUTs.[39] Those results concern a different transducer and cannot be assigned to the proposed optical microphone. An optical sensing head would require its own temperature-dependent mechanical, optical, and package measurements.

# VII.CONCLUSIONS

The model links diaphragm bending and waveguide elongation to the noise and distortion limits of direct MZI intensity readout. For the compact MEMS case, the favorable calculation gives 78 dB SNR and 124 dB AOP; the higher-power measurement case primarily improves pressure resolution. Neither comparison establishes a substantial overall performance advantage over the selected conventional devices. The main outcome is therefore a set of conditional performance estimates and design tradeoffs, together with identifiable tests of the mechanics, readout noise, overload behavior, and bandwidth. Remote optical interrogation remains a possible route to harsh-environment sensing, for which no temperature limit has yet been demonstrated by this analysis.

**Language editing disclosure.** Generative AI language tools were used to assist with English-language editing and manuscript organization.

# VIII.REFERENCES